\documentclass[9pt,twocolumn,twoside]{osajnl}

\journal{ol}

\setboolean{shortarticle}{true} 

\usepackage{amsmath}
\usepackage{amsfonts}
\usepackage{graphicx}
\usepackage{mathtools}
\usepackage{braket}

\newcommand{\diff}[2]{\frac{d #1}{d #2}}

\newcommand{\avg}[1]{\langle#1\rangle}

\newcommand{\bk}[1]{\left(#1\right)}
\newcommand{\Bk}[1]{\left[#1\right]}

\title{Quantum limits on the time-bandwidth product of an optical
  resonator}

\author[1,2,*]{Mankei Tsang}
\affil[1]{Department of Electrical and Computer Engineering,
  National University of Singapore, 4 Engineering Drive 3, Singapore
  117583}

\affil[2]{Department of Physics, National University of Singapore,
  2 Science Drive 3, Singapore 117551}

\affil[*]{mankei@nus.edu.sg}

\ociscodes{(270.2500) Fluctuations, relaxations, and noise; (230.5750) Resonators
}

\begin{abstract}
  A thought-provoking proposal by Tsakmakidis \textit{et al.}~[Science
  \textbf{356}, 1260 (2017)] suggests that nonreciprocal optics can
  break a time-bandwidth limit to passive resonators.  Here I quantize
  their resonator model and show that quantum mechanics does impose a
  limit, or else requires extra noise to be added in the same fashion
  as amplified spontaneous emission in an active resonator.  I also
  use thermodynamics to argue that extra dissipation or noise must be
  present in their proposed device.
\end{abstract}

\setboolean{displaycopyright}{true}

\begin{document}
\maketitle

Nonreciprocal optics has enjoyed a renaissance in recent years
\cite{raghu08,wang08,wang09,hafezi13,koch10,metelmann15,fang17,tian17}. While
the main driver has been the practical desire to make on-chip
isolators and circulators for integrated photonics, a recent proposal
by Tsakmakidis \textit{et al.}\ \cite{tsakmakidis17} suggests that
nonreciprocal optics can break a fundamental time-bandwidth limit to
passive resonators. If true, the proposal promises an upheaval of
fundamental physics, as well as new functionalities such as broadband
slow light. Their models and simulations are based exclusively on
classical electromagnetism, however, leaving open the question whether
they may violate other fundamental laws of physics, especially quantum
mechanics and thermodynamics.

In this Letter, I quantize their resonator model using textbook
quantum optics \cite{haus,gardiner_zoller,mandel} and show that,
unlike classical electromagnetism, quantum mechanics does impose a
time-bandwidth limit to passive resonators. To break the limit, the
quantum theory requires extra noise to be added in the same fashion as
amplified spontaneous emission in an active resonator.  Going deeper
into the details of their proposed implementation, I also find that it
requires a one-way power-transfer mechanism that violates the second
law of thermodynamics. These findings suggest that their proposal is
unlikely to be physical, or can be accomplished by a conventional
active resonator if the extra noise is acceptable.

To set the stage, I first recall basic facts regarding passive linear
optics and nonreciprocity in quantum optics. The quantum Hamiltonian
for any multi-mode passive linear optics has the general form
\begin{align}
H = \sum_{j} b_j^\dagger G_{jk} b_k,
\label{H}
\end{align}
where $\{b_j\}$ is a set of annihilation operators for the optical
modes, $\dagger$ denotes the Hermitian conjugate, and $G$ is a
mode-coupling matrix \cite{braunstein_rmp}. Since $H$ must be an
Hermitian operator, $G$ must an Hermitian matrix
($G_{jk} = G_{kj}^*$).  If I identify the $j = 0$ mode as the
resonator mode with $b_0 = a$ and the rest as reservoir modes, the
Hamiltonian becomes
$H = G_{00} a^\dagger a + \sum_{k\neq 0} (G_{0k} a^\dagger b_k +
G_{0k}^* a b_k^\dagger) + \sum_{j\neq 0,k\neq 0} b_j^\dagger G_{jk}
b_k$.  Denote the submatrix $G_{jk}$ for $j \neq 0$ and $k \neq 0$ by
$G'$. It is also Hermitian, so it can be diagonalized as
$G' = V^\dagger D V$, where $V$ is unitary, $D$ is diagonal and real,
and $\dagger$ also denotes the conjugate transpose of a matrix.
Defining the reservoir eigenmodes via
$c_j = \sum_{k\neq 0} V_{jk} b_k$ leads to the Hamiltonian
\begin{align}
H = G_{00} a^\dagger a + \sum_{j} \bk{\eta_j a^\dagger c_j +
\eta_j^* a c_j^\dagger} + \sum_{j} D_{jj} c_j^\dagger c_j,
\label{Hstart}
\end{align}
where $\eta_j \equiv \sum_{k\neq 0} G_{0k} V_{jk}^*$. This Hamiltonian
is the starting point for deriving the quantum Langevin equation for a
lossy resonator in the Markovian limit
\cite{haus,gardiner_zoller,mandel}.

Nonreciprocity in classical optics is commonly regarded as an analogy
of time-reversal-symmetry breaking in quantum mechanics
\cite{raghu08,wang08,wang09,hafezi13}. To go beyond mere analogy and
define time-reversal symmetry rigorously in quantum optics, consider
the proposal by Koch \textit{et al.}~\cite{koch10}. Let the
anti-unitary time-reversal operator be $\Theta$.  Time-reversal
symmetry in general is defined by the condition
$\Theta H \Theta^{-1} = H$ \cite{weinberg}. If $\{b_j\}$ corresponds
to a set of spatially localized modes, Koch \textit{et al.}'s
definition of $\Theta$ leads to
$\Theta b_j^\dagger \Theta^{-1} = \exp(i\theta_j)b_j^\dagger$ and
$\Theta b_j \Theta^{-1} = \exp(-i\theta_j)b_j$, where $\theta_j$ is a
phase that represents a choice of gauge. Imposing the symmetry on
Eq.~(\ref{H}),
\begin{align}
\Theta H \Theta^{-1} = \sum_{j} b_j^\dagger \exp\bk{i\theta_j-i\theta_k}G_{jk}^* b_k
= \sum_{j} b_j^\dagger G_{jk} b_k,
\end{align}
which implies
\begin{align}
\exp\bk{i\theta_j-i\theta_k}G_{jk}^* = G_{jk}.
\label{sym_Omega}
\end{align}
A $G$ matrix that cannot satisfy Eq.~(\ref{sym_Omega}) for any
$\{\theta_j\}$ breaks the time-reversal symmetry; the optical
circulator is an example \cite{koch10}. For the purpose of this
Letter, Eq.~(\ref{sym_Omega}) is simply an additional constraint on
$G$ and irrelevant to the validity of Eq.~(\ref{Hstart}), so I can
proceed with the standard formalism of quantum Langevin equations that
originate from Eq.~(\ref{Hstart}).

The simplest quantum Langevin equation for a lossy resonator is given
by \cite{haus,gardiner_zoller,mandel}
\begin{align}
\diff{a(t)}{t} &= -i\omega_0 a(t) -\frac{1}{\tau} a(t)
+\sqrt{\gamma_0} A_0(t),
\label{model1}
\end{align}
where $a(t)$ is the annihilation operator of the resonator mode in the
Heisenberg picture, $A_0(t)$ is the annihilation operator of a field
at the input of the resonator, $\omega_0$ is the resonance frequency,
$\tau$ is the decay time of the resonator mode, and $\gamma_0 \ge 0$
is the input coupling rate. Taking the expectation of
Eq.~(\ref{model1}) and substituting
$\alpha(t) = \sqrt{\gamma_0\hbar\Omega}\avg{a^\dagger(t)}$ and
$s_{\rm in}(t) = \sqrt{\hbar\Omega}\avg{A_0^\dagger(t)}$, where
$\Omega$ is the center frequency of the input field, one obtains
\begin{align}
\diff{\alpha(t)}{t} &= i\omega_0 \alpha(t) -\frac{1}{\tau}\alpha(t)
+\gamma_0 s_{\rm in}(t),
\label{classical}
\end{align}
which is the classical equation assumed by Tsakmakidis \textit{et
  al.}\ (Eq.~(3) in Ref.~\cite{tsakmakidis17}). They defined the
time-bandwidth product (TBP) as $\tau\gamma_0$ and claimed that their
proposed device could make $\tau\gamma_0$ arbitrarily large.

Assuming the standard bosonic commutation relations
$[a(0),a^\dagger(0)] = 1$ and
$[A_0(t),A_0^\dagger(t')] = \delta(t-t')$, it can be shown that
\begin{align}
\Bk{a(t),a^\dagger(t)} &=  e^{-2t/\tau}+\frac{\tau\gamma_0}{2} \bk{1-e^{-2t/\tau}}.
\end{align}
Since the Heisenberg-picture operator $a(t)$ in a closed system
evolves as $a(t) = U^\dagger(t) a(0) U(t)$, where $U(t)$ is a unitary
operator, the commutator
\begin{align}
\Bk{a(t),a^\dagger(t)} = 1
\label{commutator}
\end{align}
must be preserved at all times. The commutator must hold even for open
quantum systems, as it is well known that any open quantum system can
be modeled as part of a closed system \cite{wiseman_milburn}.  The
commutator preservation forces the TBP to be
\cite{haus,gardiner_zoller,mandel}
\begin{align}
\tau\gamma_0 &= 2.
\end{align}
In other words, the decay of the commutator at a rate of $2/\tau$ must
be compensated exactly by coupling to a quantum field at a rate of
$\gamma_0 = 2/\tau$. This is a quantum version of the
fluctuation-dissipation relation.

Coupling the resonator to more fields via passive linear optics can
only make the total dissipation rate $2/\tau$ larger and the TBP
smaller. A general model is
\begin{align}
\diff{a(t)}{t} &= -i\omega_0 a(t) - \frac{1}{\tau} a(t)
+\sum_j \sqrt{\gamma_j} A_j(t),
\label{model2}
\end{align}
where $\{A_j(t); j = 0,1,\dots\}$ are the annihilation operators of
the input fields, obeying
$[A_j(t),A_k^\dagger(t')] = \delta_{jk}\delta(t-t')$, and
$\{\gamma_j \ge 0\}$ are the coupling rates. To retrieve the classical
Eq.~(\ref{classical}), the mean field $\avg{A_j(t)}$ for $j \neq 0$
should be zero. Equation~(\ref{commutator}) demands
\begin{align}
\tau\sum_j\gamma_j = 2,
\end{align}
meaning that 
\begin{align}
  \tau\gamma_0 =2-\tau\sum_{j\neq 0}\gamma_j \le 2.
\end{align}

\begin{figure}[htbp!]
\centerline{\includegraphics[width=0.48\textwidth]{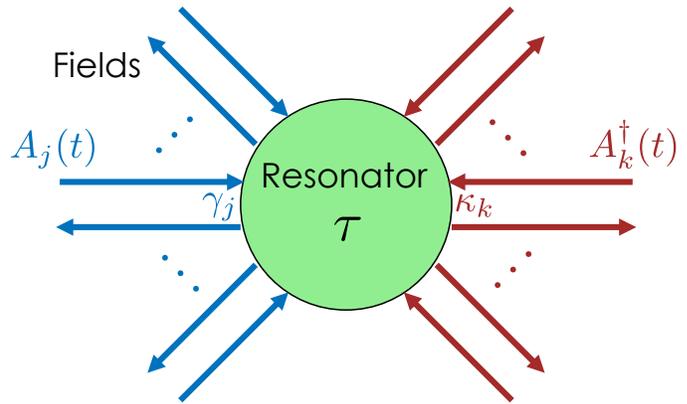}}
\caption{The general resonator model governed by Eq.~(\ref{model3}).
  Quantum mechanics demands the decay time $\tau$ and the coupling
  rates $\{\gamma_j,\kappa_k\}$ to be related by Eq.~(\ref{qtbp}).}
\label{resonator}
\end{figure}

$\tau\gamma_0 > 2$ implies that there is more quantum fluctuation from
an input than dissipation.  To keep the system linear and the
classical correspondence intact, any modification of
Eq.~(\ref{model2}) should be additive terms with zero mean, and to
make $\tau\gamma_0 >2$, the net commutator of these additive terms
should be negative to preserve Eq.~(\ref{commutator}). These
requirements can be satisfied by introducing a ``Bogoliubov'' coupling
of $a(t)$ to the creation operators
$\{A_k^\dagger(t); k = 1,2,\dots\}$ (excluding $A_0^\dagger(t)$, which
has nonzero mean), as depicted in Fig.~\ref{resonator} and described
by
\begin{align}
\diff{a(t)}{t} &= -i\omega_0 a(t) -\frac{1}{\tau} a(t)
+\sum_j \sqrt{\gamma_j} A_j(t) 
\nonumber\\&\quad
+\sum_k \sqrt{\kappa_k} e^{i\phi_k}A_k^\dagger(t).
\label{model3}
\end{align}
where $\{\kappa_k \ge 0\}$ are another set of coupling rates and the
phase factors $\{\phi_k\}$ are introduced for generality.  The
commutator of $a(t)$ becomes
\begin{align}
\Bk{a(t),a^\dagger(t)} &=  e^{-2t/\tau}+
\frac{\tau(\sum_j\gamma_j-\sum_k \kappa_k)}{2} \bk{1-e^{-2t/\tau}}.
\end{align}
Equation~(\ref{commutator}) holds when
\begin{align}
\tau\bk{\sum_j\gamma_j-\sum_k \kappa_k} &= 2,
\label{qtbp}
\end{align}
and now
$\tau\gamma_0 = 2 - \tau\sum_{j\neq 0}\gamma_j + \tau \sum_k \kappa_k$
can exceed $2$ and become arbitrarily large.  In the limit of
$\tau\to\infty$ and $\sum_j\gamma_j =\sum_k\kappa_k$, the total input
$\sum_j\sqrt{\gamma_j}A_j(t)+ \sum_k\sqrt{\kappa_k}e^{i\phi_k}
A_k^\dagger(t)$ has commuting quadratures, forming a
quantum-mechanics-free subsystem \cite{qmfs,gough} that does not
contribute to the commutator of $a(t)$.

It is well established \cite{haus,gardiner_zoller} that such a model
is equivalent to a resonator with internal amplifiers, in which case
$\kappa_k$ is the gain coefficient introduced by each amplifier and
the net dissipation rate $2/\tau= \sum_j\gamma_j-\sum_k \kappa_k$ is
reduced. The presence of $A_k^\dagger(t)$ terms in
Eq.~(\ref{model3}), however, must introduce extra noise
\cite{haus,gardiner_zoller}, known as amplified spontaneous emission
in the context of amplifiers.

When applied to Ref.~\cite{tsakmakidis17}, the quantum models
presented here imply one of the following conclusions:
\begin{enumerate}
\item If the physical device proposed by Tsakmakidis \textit{et al.}\
  obeys the passive model governed by Eq.~(\ref{model2}), the TBP
  cannot exceed 2, meaning that their classical electromagnetic model
  and simulations are unphysical. A likely explanation is that they
  failed to account for all dissipation processes in a real device.

\item If the device obeys the general model governed by
  Eq.~(\ref{model3}), it is equivalent to an active resonator and must
  suffer from extra noise to achieve $\tau\gamma_0 > 2$.  There is
  little practical need for the device when active resonators, an
  established technology, can do the same job \cite{ramezani12}. In
  fact, active resonators can offer more capabilities via quantum
  engineering
  \cite{hammerer,qnc,qmfs,gough,koch10,moeller17,metelmann14,metelmann15,fang17,tian17,wiseman_milburn}
  if the fields are accessible.

\item If the device can achieve $\tau\gamma_0 > 2$ without the extra
  noise, new physics beyond the textbook models here would be required
  to explain it. That would be a groundbreaking discovery in quantum
  optics. A non-Markovian model might offer such a possibility,
  although the classical correspondence with the Markovian
  Eq.~(\ref{classical}) would likely be lost and the definition of the
  TBP would need to be modified.
\end{enumerate}

\begin{figure}[htbp!]
\centerline{\includegraphics[width=0.4\textwidth]{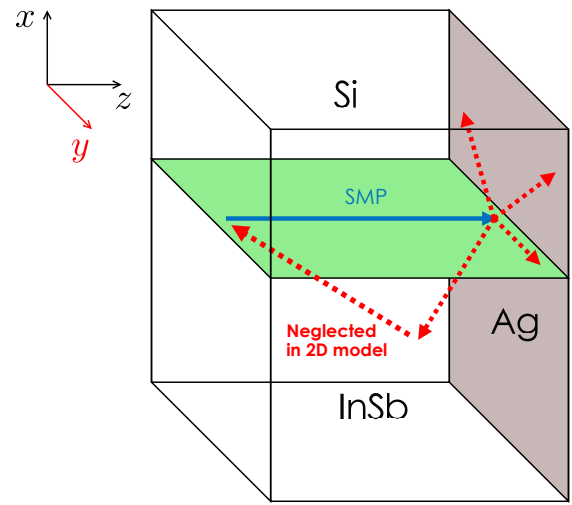}}
\caption{A three-dimensional view of the
  silicon-indium-antimonide-silver (Si-InSb-Ag) system assumed in
  Ref.~\cite{tsakmakidis17}.  Their two-dimensional model with only
  the $x$ and $z$ dimensions neglect modes that have a nonzero
  wavevector component in the $y$ dimension ($k_y$) and would miss any
  coupling of the SMPs at the Ag mirror to those modes.}
\label{corners}
\end{figure}

In the context of the nonreciprocal surface magnetoplasmons (SMPs)
considered by Tsakmakidis \textit{et al.}, one possible source of
extra dissipation is coupling to modes in a neglected third dimension,
as explained by Fig.~\ref{corners}. Their two-dimensional device model
and simulations forbid any mode that has a nonzero wavevector
component in that dimension and would therefore miss any coupling of
the SMPs to those modes in reality. Such turning at a boundary is the
more usual behavior of nonreciprocal waves
\cite{raghu08,wang08,wang09,hafezi13}.  There are two ways of modeling
this extra coupling: it can be modeled as extra dissipation for the
original resonator mode, in which case $\tau$ will decrease, or the
leaked waves can be modeled as part of a redefined resonator
supermode, in which case $\gamma$ will decrease, since the rate at
which the input can fill the spatially bigger supermode will
decrease. Either way, the TBP should decrease from their simulated
value.

\begin{figure}[htbp!]
\centerline{\includegraphics[width=0.4\textwidth]{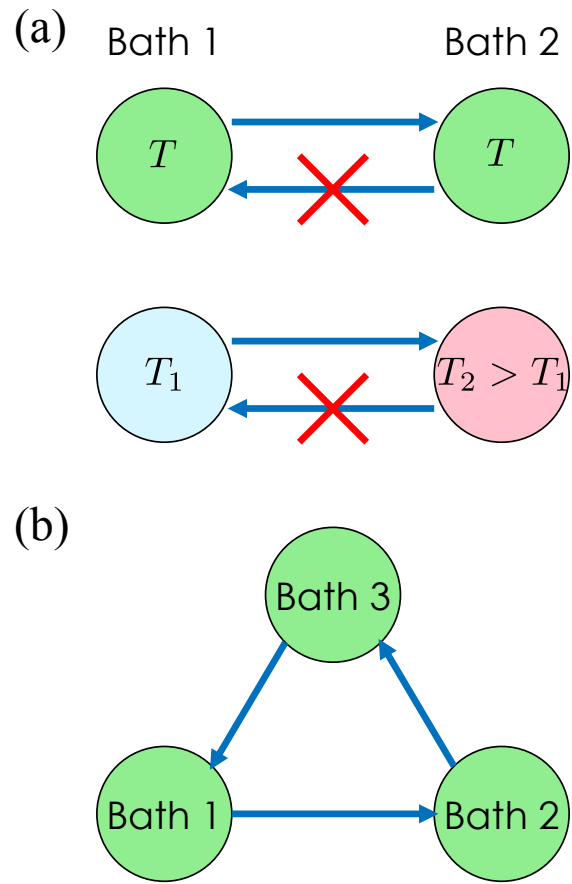}}
\caption{(a) Two equal-temperature baths undergoing one-way power
  transfer would develop a temperature difference, violating the
  second law of thermodynamics. (b) A third bath is needed to restore
  the second law in a closed nonreciprocal system.}
\label{thermo}
\end{figure}

Thermodynamics can provide another argument that extra dissipation or
noise must exist.  The model in Ref.~\cite{tsakmakidis17} relies
crucially on the assumption that only a forward-propagating SMP mode
exists and the back-propagating mode is forbidden, resulting in a net
power transfer and energy build-up in one direction. Suppose that two
baths, initially at the same temperature, are placed on the opposite
ends of the system, exchanging energy via the SMPs only. The net
one-way power transfer would lead to a temperature difference between
the two baths, violating the second law of thermodynamics, as depicted
in Fig.~\ref{thermo}(a). To restore the second law in a closed system,
there must be coupling to a third bath that is neglected in their
model, as depicted in Fig.~\ref{thermo}(b).  Whether this coupling to
the third bath introduces extra dissipation---thereby reducing the
TBP---or extra noise depends on the nature of the coupling.

It is interesting to note that Lord Rayleigh recognized the apparent
conflict between nonreciprocal optics and thermodynamics as early as
1885 \cite{rayleigh85}. Wien proposed a magneto-optical scheme to
achieve one-way energy transfer between two ports and violate the
second law \cite{wien00}, but Rayleigh realized that Wien's scheme
actually involves more than two ports, and closing those extra ports
via mirrors makes the two-port coupling rates identical
\cite{rayleigh01}. Similarly, to model a passive two-port optical
isolator that absorbs the back-propagating light, a third port that
takes the absorbed light should be included, and the model becomes
equivalent to a three-port circulator.  The circulator input-output
relations, though nonreciprocal, remain unitary \cite{garrison} and
thus compliant with thermodynamics.  Even in the context of SMPs, it
has been recognized that thermodynamics mandates nonreciprocal
absorption \cite{remer} or more than two ports \cite{stamps} for
nonreciprocal reflections to occur. Another interesting example is the
chiral edge modes that arise from time-reversal-symmetry breaking in
quantum Hall systems as well as photonics
\cite{raghu08,wang08,wang09,hafezi13}. The modes can propagate in only
one direction along the edge of a two-dimensional surface and may
sound at odds with thermodynamics, but note that the edge of a finite
surface is always a closed curve. Owing to the breaking of
time-reversal symmetry, the path that transfers energy from Bath 1 at
one point to Bath 2 at another point is not the time reversal of the
path that transfers energy from Bath 2 to Bath 1.  Both paths do exist
on a closed curve, however, and thermodynamics is safe. All these
examples show that the modeling of passive nonreciprocal optics
requires utmost care, to ensure compliance with not only classical
electromagnetism but also thermodynamics.

Regardless of the implementation, the nonreciprocal resonator
hypothesized in Ref.~\cite{tsakmakidis17}, with its large input rate
and arbitrarily small output rate, is analogous to a black hole that
sucks energy in one direction.  If there is no other neglected
dissipation, extra noise akin to Hawking radiation must be present to
uphold the second law.  The active-resonator model offers a natural
way to extend the analogy to the quantum domain: the type of
Bogoliubov coupling used in Eq.~(\ref{model3}) also underpins the
quantum theory of black-hole radiation \cite{hawking74} and ensures
consistency with both quantum mechanics and thermodynamics.


\textbf{Funding.} Singapore Ministry of Education Academic Research
Fund Tier 1 Project R-263-000-C06-112.

Interesting discussions with Kosmas Tsakmakidis, who brought
Refs.~\cite{ramezani12,raghu08,wang08,wang09} to my attention, are gratefully
acknowledged.  

\bibliography{time-bandwidth2}
\bibliographyfullrefs{time-bandwidth2}

\end{document}